# A 3D Memristor Architecture for In-Memory Computing Demonstrated with SHA3

Muayad J. Aljafar, Rasika Joshi, and John M. Acken

**Abstract**—Security is a growing problem that needs hardware support. Memristors provide an alternative technology for hardware-supported security implementation. This paper presents a specific technique that utilizes the benefits of hybrid CMOS-memristors technology demonstrated with SHA3 over implementations that use only memristor technology. In the proposed technique, SHA3 is implemented in a set of perpendicular crossbar arrays structured to facilitate logic implementation and circular bit rotation (Rho operation), which is perhaps the most complex operation in SHA3 when carried out in memristor arrays. The Rho operation itself is implemented with CMOS multiplexers (MUXs). The proposed accelerator is standby power-free and circumvents the memory access bottleneck in conventional computers. In addition, our design obscures the intermediate values from the I/O interface and outperforms the state-of-the-art memristor-based designs in terms of size and energy. Demonstrating the memristor implementation of SHA3 provides an impetus for utilizing memristors in information security applications.

## 1 INTRODUCTION

CRYPTOGRAPHIC hash functions are widely used in information security applications such as digital signature generation and verification, pseudorandom bit generation, and key derivation. The need for high-speed secure implementation of such functions is generally true, but mainly for big data applications. Software implementations of hash functions fall short in performance and security; therefore, we focus on hardware implementations. Various CMOS-based hardware accelerators for the recent standard secure hash algorithm, SHA3, have been proposed. Some of these accelerators were optimized for the area [1] speed [2] reliability [3], and throughput [4]. Yet, the distance between memory and logic units in the accelerators results in additional power dissipation and delay when reading and writing the data on top of performing the hash algorithm. At deep sub-100nm fabrication nodes, the leakage power and data traffic are significant. When hashing large objects stored in memory, data transfer becomes a bottleneck. To overcome the bandwidth limitation of data bus traffic, this paper relies on memristor technology for in-memory computing—a computing model where *calculations* are implemented in *memory*, unlike the conventional computing model where the calculation unit is separate from the memory unit. The benefit of in-memory computing is well described in [5], [6]. Computations in this computing paradigm, however, are sequential and slow. We propose a new technique that utilizes the benefits of a hybrid CMOS-memristors technology demonstrated with SHA3 over implementations that use only memristor technology. Our accelerator is highly optimized for area and energy and can be used in resource-constrained applications. The details of the comparisons are shown in Section 5.

**Previous Work**: Emerging technologies have been employed in the implementation of SHA3. Oved et al. [7], for instance, utilized a general-purpose in-memory computing architecture to implement SHA3. They additionally devised a technique for the bit rotation operation and optimized their design for energy efficiency. Bhattacharjee et al. [8] employed ReRAM technology to compute the round function in SHA3 within resource-constrained Internet of Things (IoT) nodes. This computation was realized in ReRAM-based VLIW architecture (ReVAMP), comprising two 1S1R crossbar arrays and lightweight peripheral circuitry. Notably, one of the crossbar arrays was dedicated to instruction memory, while the other was utilized for data storage and computation. The implementation of SHA3 involved a set of majority functions and inversion operations. Xue et al. [9] introduced a memristive RISC-V processor designed to facilitate in-memory computing for blockchain technology. This processor was specifically developed for Internet of Things (IoT) applications. The authors demonstrated a general compiling policy for implementing the SHA3 algorithm as an illustrative example. Yang et al. [10] carried out the implementation of SHA3 using a crossbar array of voltage-gated spin hall-effect-driven magnetic tunnel junctions (VG-MTJ). Their design included both a non-pipelined single message hash circuit

―――――――――――――――

- M.J. Aljafar is with Xenergic AB, Scheelevägen 15, 223 70 Lund, Sweden. E-mail: muayad.aljafar@xenergic.com.
- R. Joshi is with Intel Hillsboro, OR, USA. E-mail: joshiras@pdx.edu.
- J.M. Acken is with the Electrical and Computer Engineering Department, Portland State University, Portland, OR 97201. E-mail: acken@pdx.edu.

and a pipelined multiple message hash circuit to execute SHA3, incorporating stateful logic gates. Additionally, Nagarajan et al. [11] put forward a high-performance, area-efficient implementation for SHA3 named SHINE, which is based on ReRAM technology. SHINE facilitates in-memory computing and executes various functions in a sum-of-products (SOP) form within crossbar arrays.

**The proposed work:**

We present a 3D CMOS-memristor accelerator designed for SHA3 implementation. The accelerator consists of four sets of memristor arrays: *state*, *Rho*, *Chi*, and *Iota*. These arrays perform distinct functions and communicate through CMOS transmission gates. The state array functions both as a logic and memory unit, while the Rho array exclusively operates as a memory unit. Additionally, the Chi and Iota arrays are dedicated to logic and memory operations, respectively. Structured for SHA3 execution within a compact architecture, these arrays directly incorporate the data representation outlined in Section 4, facilitating parallel and pipeline computations. The memristor arrays utilize rectifying memristors, specifically those with diode behavior. Opting for crossbar arrays with rectifying memristors, instead of non-rectifying ones, offers the advantage of suppressing sneak currents and simplifying the memristor arrays by eliminating the need for external selectors. Furthermore, the use of rectifying memristors enables the implementation of native volistor gates, such as XOR/XNOR, seamlessly integrated into our design.

The accelerator exhibits high speed, with its primary focus on optimizing area and energy overheads. It capitalizes on the benefits of hybrid CMOS-memristor technology, specifically integrating CMOS in the implementation of a circular bit rotation operation, known as the Rho operation. This strategic use of CMOS simplifies the design, especially in managing the complexities associated with executing the Rho operation solely with memristor arrays. Notably, the Rho operation in our accelerator is completed in just five clock cycles. Alternatively, achieving a high-speed implementation of the Rho operation solely using memristor arrays would necessitate extensive parallel operations, resulting in significant area overhead (Section 5). The proposed accelerator achieves a well-balanced tradeoff among energy, area, and speed, attributed to our architectural decisions and the utilization of rectifying memristors.

In summary, this work introduces a method for implementing the SHA3 algorithm using perpendicular crossbar arrays. These arrays and their interconnections are designed to support logic implementation and the circular bit rotation operation, which is arguably the most intricate operation when implemented in memristor technology. The circuit structure employed, as detailed in the initial paragraph of Section 4, constitutes the primary innovation of this work.

**Fabrication Feasibility**: Advancements in 3D memristor fabrication technology affirm the feasibility of our architecture [12]. For instance, the procedures detailed in [13] showcase the fabrication of multiple stacked crossbar arrays using rectifying memristors, aligning with the memristor type employed in our design. Moreover, the idea of arranging rotated crossbar arrays alongside each other in our design has been extensively deliberated in various publications [14], [15], [16], [17]. These examples in the literature underscore the viability of constructing 3D crossbar arrays akin to our proposed design.

The suitability of memristors for cryptography lacks the same tolerance observed in neuromorphic applications. Testing memristors is beyond the scope of our paper and has been extensively covered in numerous publications [18], [19]. Moreover, memristors are still emerging devices, and consequently, attack models in this technology, unlike CMOS, have not been extensively developed. While this presents an advantage in applying memristors to security applications, there is a need for more information on the reliability and fault tolerance guidance of memristors.

The security analysis of the proposed circuit falls beyond the scope of this paper; however, the performance evaluation of the circuit is provided in Section 5. It is noteworthy that memristor technology has been applied in security applications, such as physically unclonable functions (PUF) [20], [21], and true random number generators (TRNG) [22], where the stochastic switching behavior in memristors was exploited as a source of entropy. In this work, the stochastic behavior of memristors is not utilized; rather, the memristors are employed for logic and memory operations.

The remainder of the paper is structured as follows. Sections 2 and 3 provide a concise review of memristor technology and the SHA3 standard, respectively. Sections 4 and 5 present the introduction and analysis of our SHA3 implementation. Finally, Section 6 serves as the conclusion of the paper.

## 2 BACKGROUND ON MEMRISTORS

The existence of memristors as the fourth circuit element was mathematically predicted by Chua in 1971 [23] and in 2008, a group of researchers at HP Labs showed the first analytical example of a memristor [24]. Memristors are non-volatile nanoscale devices that enable in-memory computing [25], [26]. A common structure for memristive circuits is crossbar arrays. They support a large number of connections in a small footprint, however, crosstalk due to sneak paths of current limits the size of these crossbar arrays. There are techniques for mitigating the sneak paths in crossbar arrays [27], [28]. In this work, we rely on rectifying memristors [29], [30], [31] to reduce the sneak paths effect. Specifically, the rectifying memristors have a large $R_{OFF}/R_{ON}$ ratio (e.g., $10^3 \leq R_{OFF}/R_{ON} \leq 10^6$) that aids to mitigate the effect of the sneak paths in crossbar arrays and leaves a margin to differentiate between high and low resistance states, hence improving logic computations. In addition, this type of memristors enables the implementation of *volistor* and *programmable diode gates*—styles of memristor logic gates utilized in this paper for implementing the SHA3 algorithm [26], [32], [33]. Fig. 1 shows the *i-v* characteristic and symbolic diagram of a rectifying memristor, illustrating the response to an applied sinusoidal signal with an amplitude of 1.2V and a frequency of 100MHz. The diagram has been replicated from [39], employing the LTspice simulator. Programming the memristor to low resistance state $R_{ON}$ requires applying $V_{SET}$ across the device. $V_{SET}$ is a positive voltage larger than the voltage threshold $V_{CLOSED}$. Similarly, programming the memristor to high resistance state $R_{OFF}$ requires applying $V_{CLEAR}$ across the device. $V_{CLEAR}$ is a negative voltage smaller than the voltage threshold $V_{OPEN}$. Applying a voltage between $V_{OPEN}$ and $V_{CLOSED}$ will not change the current state of a memristor. The equation (1) defines the resistance and dynamics of the memristor, as outlined in [39]. In this equation, $R$ represents the resistance of the memristor, and $v$ is the applied voltage across the memristor. The memristor is described by a state variable $w \in [0, 1]$, representing its resistance in the forward-biased direction. Specifically, when $w=0$, the memristor is in the High-Resistance State (HRS), exhibiting $R_{OFF}$ resistance. Conversely, when $w=1$, the memristor is in the Low-Resistance State (LRS), demonstrating $R_{ON}$ resistance. The positive constant $\alpha$, representing the programming rate, is assumed to be $1 \times 10^9$ (Vs)$^{-1}$. With this specified value of $\alpha$, a memristor initially in state $w=0$ can be programmed to state $w=1$ in 1ns by applying a voltage of +1.2V across it. The parameter values employed for the memristor model align with those described in [39].

$$R = \begin{cases} R_{OFF} \left(\frac{R_{ON}}{R_{OFF}}\right)^w, & v \geq 0 \\ R_{OFF}, & v < 0 \end{cases} \quad (1)$$

$$\frac{dw}{dt} = \begin{cases} \alpha(v - v_{CLOSED}), & v \geq v_{CLOSED} \\ \alpha(v - v_{OPEN}), & v \leq v_{OPEN} \\ 0, & v_{OPEN} < v < v_{CLOSED} \end{cases}$$

In this study, the logic '1' state is represented by $R_{ON}$, and the logic '0' state is represented by $R_{OFF}$.

A concise overview of the volistor XOR/XNOR gate and the programmable diode AND/NAND gate employed in the SHA3 design is provided below. For additional information, please refer to [32], [33], [34]. Volistors possess the ability to implement in-memory computing. Their operation involves using voltage as input and resistance as output, leveraging the diode behavior of rectifying memristors. In contrast, programmable diode gates lack the capability of in-memory computing. They depend on rectifying memristors but utilize voltage as both input and output. Both volistors and diode gates are exclusively implemented in two-dimensional crossbar arrays. This is in contrast to stateful logic gates, which have the flexibility to be realized in one-dimensional memristor arrays (such as 1×$n$ or $n$×1).

The decision to use memristors for diode gates, traditionally implemented with resistors and diodes, may raise questions for readers. However, the rationale behind this choice is twofold. Firstly, crossbar arrays of rectifying memristors can execute diode gates using the same CMOS drivers employed for stateful logic gates, thereby avoiding additional complexity or area overhead in the CMOS drivers. Secondly, these gates can be cascaded with volistor logic gates within crossbar arrays to achieve subsequent logic levels. This proves to be a time-efficient solution for implementing functions suitable for sum-of-products or product-of-sums structures. This advantage cannot be solely attained through the use of stateful logic gates, as highlighted in [32].

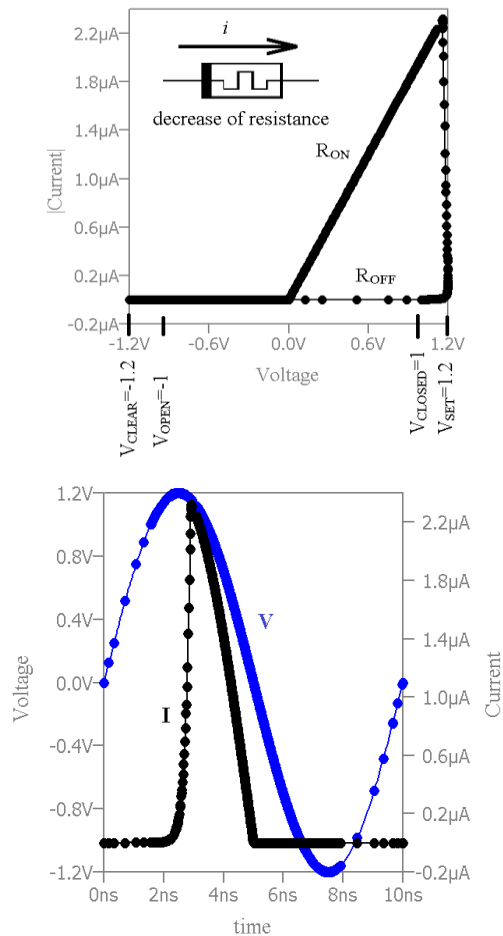

Fig. 1. $i$–$v$ characteristic of a rectifying memristor (top). The flow of current into the device leads to a decrease in resistance. Applied voltage across the memristor and the corresponding current flowing through it (bottom).

Programmable Diode AND/NAND Gate: Our accelerator employs two distinct implementations for computing logic AND/NAND, as explained below. The AND gate is utilized to read the output of a volistor XNOR gate, while the NAND gate is employed in computing the Chi operation (Section 3).

## 2.1 Diode AND Gate Implementation in XNOR Operation

The volistor XNOR gate employs a diode AND gate to generate its output. In Fig. 2a, the circuit calculates the AND gate, with memristors A and B storing inputs and memristors Y1 and Y2 determining the output ($v_{out}$). This design utilizes a crossbar array for both memory and computation. The chosen voltage levels are {0, $v^+$}, where $v^+$ is a positive voltage smaller than $V_{CLOSED}$ to avoid destructive effects on memristor states. The voltage $v^+$ is used for reading inputs, and input values ($v_A$ and $v_B$) are applied to memristors Y1 and Y2 for AND logic computation. Wire Y functions as a wired-AND connection influenced by voltage levels and the polarities of the memristors to which inputs are applied. Input voltages are applied to *terminal B* of memristors Y1 and Y2. $R_g$ is a reference resistor chosen such that $R_{ON} \ll R_g \ll R_{OFF}$. In this setup, it's crucial to program Y1 and Y2 to the LRS. Fig. 2b demonstrates the correct behavior of the AND gate when inputs '0' and '1' are stored in memristors A and B, respectively. Importantly, all memristors in the crossbar array maintain their resistance states during operation. In this specific case, memristor Y2 is reverse-biased, i.e., the voltage across Y2 is approximately - $v^+$. Consequently, even though Y2 has been programmed to the LRS, it exhibits a high-resistance state. As a result, Y2 effectively suppresses sneak current, contributing to a robust logic implementation.

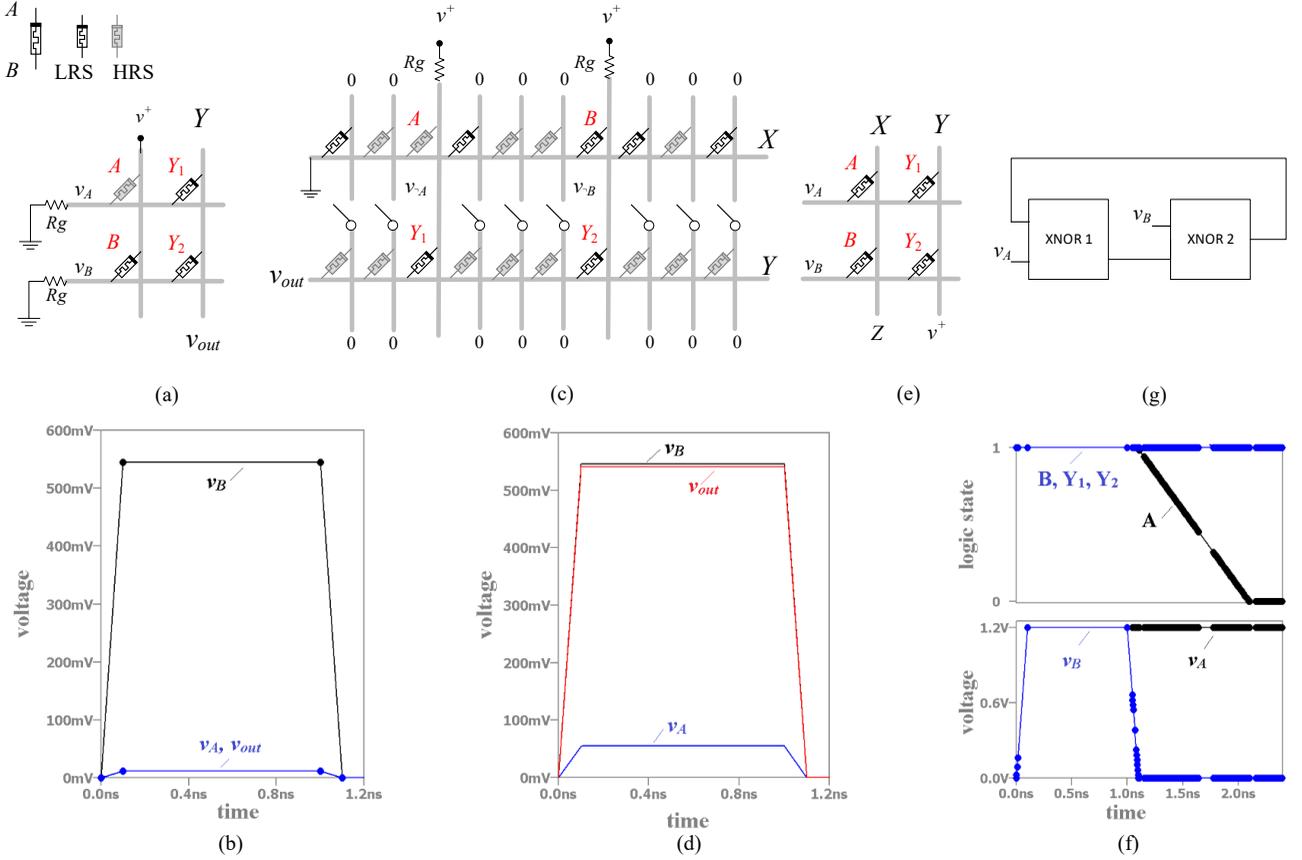

Fig. 2. Implementation examples of programmable diode AND/NAND gates and volistor XOR/XNOR gates. (a) Implementation of a diode AND gate. (b) Behavior of the diode AND gate. (c) Implementation of a diode NAND gate. (d) Behavior of the diode NAND gate. (e) Implementation of a 2-input XNOR gate. (f) Behavior of the XNOR gate. (g) Block diagram of multi-input XOR gate.

### 2.2 Diode NAND Gate Implementation in Operation Chi

The circuit depicted in Fig. 2c operates as a NAND gate. Memristors A and B serve as input storage, while Y1 and Y2 compute the output ($v_{out}$). In this configuration, CMOS transmission gates are employed to isolate the memory and calculation arrays. Similar to the diode AND gate, this circuit operates with voltage levels {0, $v^+$}. The voltage $v^+$ is utilized for reading input values, resulting in the generation of inverted inputs ($v_{\neg A}$ and $v_{\neg B}$). These inverted inputs are then directed to memristors Y1 and Y2 to implement NAND logic. Wire Y functions as a wired-OR connection, and its behavior is influenced by both voltage levels and the polarities of the memristors to which inputs are applied. The inputs are directed to terminal A of Y1 and Y2, and it is crucial to program these memristors to the LRS. Memristors on row Y that are not actively involved in the current operation have their vertical wires set to 0V. This configuration has the potential to induce a reverse-biased state in these memristors, causing them to exhibit a high-resistance state without requiring explicit programming. This setup effectively suppresses sneak current, ensuring a robust logic implementation. Fig. 2d illustrates the correct behavior of the NAND gate with '0' and '1' inputs in memristors A and B, respectively. All memristors in the crossbar array maintain their resistance states. In this scenario, Y2 is reverse-biased, suppressing sneak current and contributing to a robust logic implementation.

## 2.3 2-Input Volistor XNOR Gate Implementation

2-input volistor XOR gates are utilized in the Chi and Iota operations. The setup illustrated in Fig. 2e is employed for implementing the XOR gate. It functions by applying inputs $v_A$ and $v_B$ to memristors A and B, and the output is generated through the computation of AND logic based on the state values of memristors A and B. This computation is carried out by memristors Y1 and Y2. In this particular design, input storage and calculation arrays are isolated through CMOS transmission gates. The chosen input voltage levels are {$|V_{CLEAR}|$, 0} with $V_{CLEAR}$ representing a destructive voltage that affects the states of memristors (Fig. 1). In this setup, it is essential to initialize all memristors to the LRS, corresponding to logic '1'. If the inputs are the same, there will be no state transition in either memristor A or B, resulting in an output of AND (1, 1) or logic '1'. On the other hand, when the inputs are different, the state of a memristor connected to the input $|V_{CLEAR}|$ will toggle, leading to an output of AND (0, 1) or logic '0'. This behavior corresponds to the correct implementation of XNOR logic. In this configuration, wire X serves as a wired-AND connection and displays 0V when applied inputs are non-identical. In this scenario, the voltage across a memristor connected to $|V_{CLEAR}|$, which is $V_{CLEAR}$, is adequate to toggle the state of that memristor. Throughout this process, wire X is connected to high impedance Z, and wire Y is connected to $v^+$ to prevent any state transition in memristors Y1 and Y2. Fig. 2f shows the behavior of the XNOR gate for both identical and non-identical inputs. The state transition in memristor A connected to $|V_{CLEAR}|$ is evident. Additionally, Fig. 2f shows the states of Y1 and Y2, effectively protected against any unintended change. The logic AND of the state values of memristor A and B, representing the output, is calculated in the same manner as shown in Fig 2a. This specific step is not explicitly illustrated in Fig. 2f. Reusing the XOR gate necessitates programming memristors A and B to the LRS using $V_{SET}$.

## 2.4 Multi-Input Volistor XOR Gate Implementation

This gate is used to implement the Theta operation. Fig. 2g shows the block diagram of the XOR gate. Connecting the XNOR blocks, as illustrated in Fig. 2g, required adjusting the voltage levels at the blocks' outputs due to the different voltage levels used at the input and output of the XNOR gates, i.e. {0, $|V_{CLEAR}|$} and {0, $v^+$}. This adjustment is accomplished by cascading inverters with different voltage supplies.

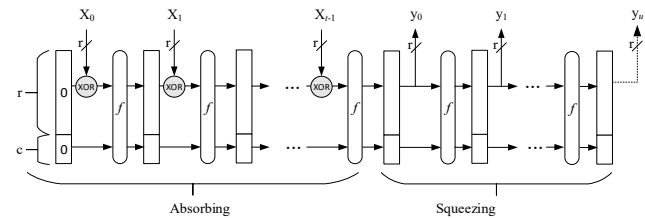

Fig. 3. The sponge construction. The structure consists of absorbing and squeezing phases [35].

## 3 SHA3 STANDARD

A cryptographic hash function is a mathematical function that takes a message and applies a set of mathematical transformations to produce a digest. SHA3 is a secure hash function and a subset of Keccak, a family of cryptographic primitives based on sponge construction [35], [36], [37]. The sponge construction consists of two phases: absorbing and squeezing (Fig. 3). In the absorbing phase, the algorithm absorbs and processes message blocks $X_i$ ($i = 0, \cdots, t-1$), and in the squeezing phase, it outputs (squeezes) the digest $y_j$ ($j = 0, \ldots, u$). Various parameters govern the dimensions of a message block, the resulting digest, and the security level of the Keccak algorithm. These parameters include the state width $b$, bit rate $r$ (representing the size of the message block), and capacity $c$, where $b = r + c$. In the standard SHA3 algorithm, $b$=1600 bits arranged in a 5×5 array of 64-bit elements as shown in Fig. 4. The array is denoted by $A[x, y]$, where $x$ and $y \in$ {0, 1, 2, 3, 4}. Function Keccak-$f$ is the heart of SHA3 and is used in both absorbing and squeezing phases. Fig. 5 illustrates the internal structure of function Keccak-$f$. The function takes as input the concatenation of ($X_i \oplus r$) and $c$ and produces the new state of the Keccak-$f$ function as output. Initially, both $r$ and $c$ are initialized to zero, therefore, $f(b) = f(X_0 || c)$ where $||$ signifies concatenation operation. Function Keccak-$f$ consists of $n_r$ rounds. In each round, a sequence of five operations—θ, ρ, π, χ, and ι—is implemented. These operations contribute to the diffusion and confusion properties necessary for security. The rounds are identical except for the operation ι, where the constants $RC[i]$ vary in each round. In the SHA3 standard, $n_r$ is set to 24. The operations are further detailed below.

**Theta (θ):** Theta is the first operation in the round sequence. It involves a linear transformation applied to the array $A[x, y]$. This transformation combines the bits of the array $A$ using a specific XOR pattern as defined by equation (2), where rot($C[x+1]$, 1) represents a 1-bit circular bit rotation operation applied to the array $C[x+1]$. This operation plays a critical role in contributing to the diffusion and confusion properties within the cryptographic SHA3 algorithm.

$$\begin{cases} A' = \theta(A): \\ C[x] = A[x,0] \oplus A[x,1] \oplus A[x,2] \oplus A[x,3] \oplus A[x,4] \\ D[x] = C[x-1] \oplus rot(C[x+1],1) \\ A'[x,y] = A[x,y] \oplus D[x] \end{cases} \quad (2)$$

**Rho (ρ) and Pi (π):** Rho and Pi operations involve a circular bitwise rotation and column permutation operations within the array $A[x, y]$. These operations are clarified by equation (3), where $r[x, y]$ denotes the rotation offset as specified in TABLE 1. As an example, the expression $A'[1, 4]$= $rot(A[3, 1], r[3, 1])$ applies a rotation offset of 55 to lane $A[3,1]$ and stores the resulting lane in $A'[1, 4]$. The term 'lane' is defined in Fig. 4.

$$\begin{cases} A' = \pi(\rho(A)) \\ A'[y, 2x+3y] = rot(A[x,y], r[x,y]) \end{cases} \quad (3)$$

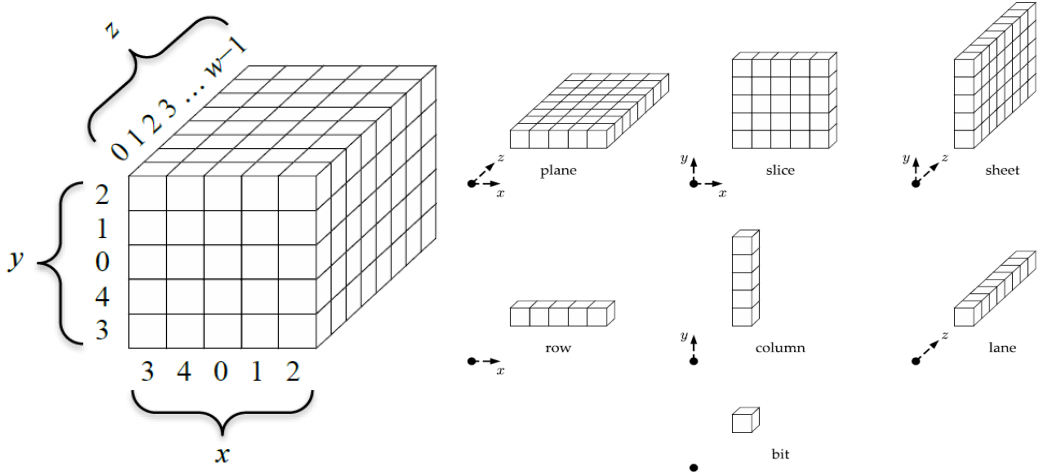

Fig. 4. Illustration of state width b and its parts organized by dimensions [36]. The numbering of the $x$ and $y$ order is from the NIST standard [38].

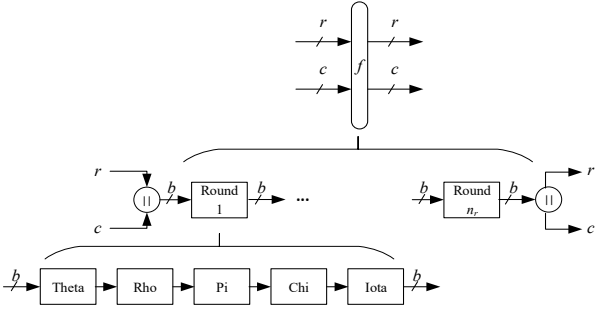

Fig. 5. Internal structure of function Keccak-$f$ [37].

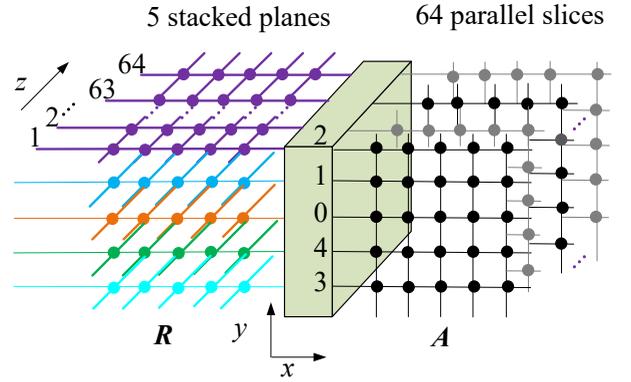

Fig. 6. Schematic of the proposed architecture (side view).

**Chi($\chi$)**: The Chi ($\chi$) operation is a non-linear transformation that involves applying a set of logical operations to the bits within each row of the array based on the values of neighboring bits. The Chi operation is expressed by equation (4), where $\neg A[i, j]$ represents the complement of $A[i, j]$, and $\wedge$ denotes logical AND. This operation contributes to introducing non-linearity and enhancing the confusion properties within the cryptographic algorithm.

$$\begin{cases} A' = \chi(A): \\ A'[x,y] = A[x,y] \oplus ((\neg A[x+1,y]) \wedge A[x+2,y]) \end{cases} \quad (4)$$

**Iota ($\iota$)**: Iota is the fifth and final operation in the round sequence, following $\theta$, $\rho$, $\pi$, and $\chi$. It involves the XOR of a specific constant denoted as $RC[i]$ with $A[0, 0]$. This operation is detailed by equation (5), where $i \in \{1, 2, ..., 24\}$. The constants $RC[i]$ play a significant role in introducing variability into the algorithm, and the specific values for each round can be found in the [37]. In all equations, (2)-(5), it is necessary to apply modulo 5 to the indices to ensure proper computation. Readers are urged to consult [37] for further insights into the operations within the standard SHA-3 algorithm. Section 4 presents a novel design implementation for these operations utilizing memristor technology.

$$\begin{cases} A' = \iota(A): \\ A'[0,0] = A[0,0] \oplus RC[i] \end{cases} \quad (5)$$

## 4 PROPOSED IMPLEMENTATION OF SHA3

Fig. 6 illustrates a logical representation of the proposed circuit for implementing SHA3. The diagram comprises two sets of perpendicular crossbar arrays connected through MUXs (enclosed in the green box). The CMOS circuits and memristors are situated on separate layers, with their physical alignment facilitated by a layer-to-layer interconnect [34]. Memristors are simplified as colored circles for clarity. On the right side, 64 *parallel slices* of 5×5 without interconnection (refer to Fig. 4 for the 'slice' definition) store and process the state values of SHA3. On the left side, five *stacked planes* of 5×64 without interconnection (refer to Fig. 4 for the 'plane' definition) serve as memory to store intermediate values. Additional components of the circuit will be elaborated later in this section.

Arranging the crossbar arrays as described above yields low overhead and high circuit performance. These arrays are optimal for organizing a message block in three dimensions, aligning directly with the SHA3 structure. The data on both sides of MUXs can be conceptualized as two 5×5 arrays (*A* and *R*) of 64-bit elements. This configuration forms a short pipeline circuit where the message can efficiently traverse from crossbar array *A* to crossbar array *R* and undergo processing in just a few clock cycles. The compact memristor crossbar arrays occupy a minimal area above the CMOS substrate. Apart from their geometric advantages, the crossbar arrays operate without standby power, resulting in zero static power consumption. The implementation of the Keccak-*f* function begins with the Theta operation. However, since the Rho operation is integral to implementing the Theta operation (i.e., *rot(A[],1)*), we will first provide details on the implementation of the Rho operation.

### 4.1 Implementation of Operation Rho

In Fig. 6, the green box is responsible for applying the Rho operation to the planes of the state array $A[x, y]$. This box is composed of five lanes, each containing 64 MUXs. Within each lane, a set of offset values, as specified in row *y* in TABLE 1 ($y \in \{0, 1, 2, 3, 4\}$) is applied to the corresponding lanes of plane *y* in the state array *A*.

In Fig. 7, a top view reveals the state array on the right connected to the Rho array on the left in the *plane y=0* through corresponding MUXs situated in *lane y=0*. Each MUX in this lane has five inputs, with each input corresponding to one offset in row *y=0* in TABLE 1. For instance, red and black wires in the illustration correspond to offset 1 and offset 0, respectively. Please note that only a subset of these wires is depicted due to size constraints.

The Rho operation is executed over five clock cycles. During each clock cycle, data in sheet *x* of the state array undergo rotation and are then stored in sheet *x* of the Rho array (refer to Fig. 4 for the 'sheet' definition). Precisely, data residing in the lanes of sheet *x* are simultaneously fed into corresponding lanes of MUXs, and the resulting outputs are stored in the Rho array. In essence, MUXs apply the offset values of column *x* in TABLE 1 to sheet *x* in the state array $A[x, y]$ within a single clock cycle. Consequently, it requires five clock cycles to apply the offset values of all five columns, completing the implementation of the Rho operation.

In this implementation, each sheet of the state array $A[x, y]$ is linked to a specific input of the MUXs. Consequently, all MUXs can function with identical control select bit values. By connecting the three control select bits of all MUXs to a specific value for each column of offsets (e.g., 001 for offsets of column *x=1* or 010 for offsets of column *x=2*), the data rotation can be executed using only three control select bits.

In the implementation of operation Theta, it is necessary to apply offset 1 to every lane of $A[x, y]$ (refer to equation 2). However, in this design, offset 1 is confined to the offset values in row *y=0* (TABLE 1), affecting solely the lanes in the plane *y=0* within the state array. As a result, the design needs to be adjusted to implement offset 1 by incorporating 6-input MUXs in the lanes of *y*={1, 2, 3, 4} within the MUXs.

### 4.2 Implementation of Operation Theta

This implementation necessitates updating the circuit from

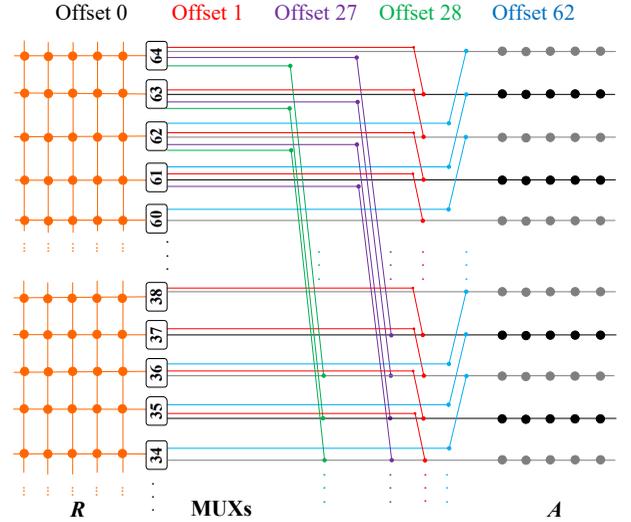

Fig. 7. The state array A is connected to the Rho array R through MUXs. Due to the size limit, only some MUXs' inputs are shown. Colors correspond to different offsets as the legend explains.

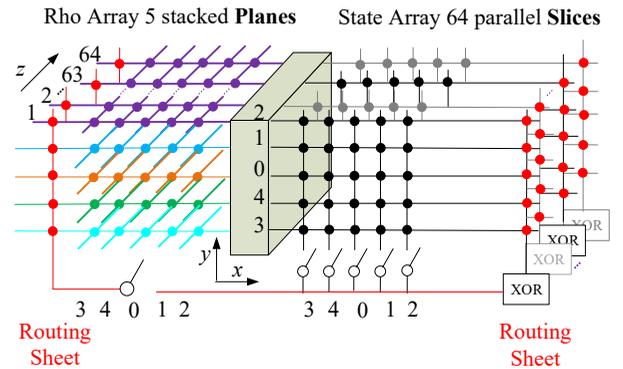

Fig. 8. Implementation of the Theta operation.

Fig. 6 to the one illustrated in Fig. 8. The revised circuit facilitates data movement from the Rho array to the state array through CMOS transmission gates. Furthermore, slices on both sides are expanded to facilitate data routing. For instance, the routing sheet on the left side establishes vertical connections between the stacked planes. Additionally, on the right side, memristors at the bottom right corner of every slice are substituted with multi-input volistor XOR gates [33]. The operational steps for executing operation Theta are clarified as follows. For $x, y \in \{0, 1, 2, 3, 4\}$,

1. Implement the Rho operation with offset 1 on all five

TABLE 1
OFFSETS OF RHO [37]

|  | x=3 | x=4 | x=0 | x=1 | x=2 |
|---|---|---|---|---|---|
| y=2 | 25 | 39 | 3 | 10 | 43 |
| y=1 | 55 | 20 | 36 | 44 | 6 |
| y=0 | 28 | 27 | 0 | 1 | 62 |
| y=4 | 56 | 14 | 18 | 2 | 61 |
| y=3 | 21 | 8 | 41 | 45 | 15 |

lanes of sheet $x+1$ in the state array and store the outputs in the corresponding sheet in the Rho array. In other words, perform the operation: $R[x+1, y]=\rho(A[x+1, y], 1)$. This step is carried out within one clock cycle per sheet.

2. Perform a bitwise XOR operation on column $x-1$ in the state array and column $x+1$ in the Rho array, involving XORing 10 bits in the specified columns. The resulting output represents the final state of the multi-input volistor XOR gate used in this implementation. Simultaneously execute the XOR operation on all 64 parallel slices of the state and Rho arrays, storing the results in the corresponding XOR gates. In other words, perform $X=X\oplus A[x-1, y]\oplus R[x+1, y]$, where $X$ is the resulting outputs of the XOR gates and $y \in \{0, 1, 2, 3, 4\}$.

    The XOR gates function sequentially, with each gate XORing a current bit with the previously calculated bit in two clock cycles. As a result, this process of calculating 64 10-input XOR gates is implemented over 19 clock cycles.

3. Perform XOR operations on the outputs of the XOR gates $X$ with a specified lane in the state array, e.g., $A[x, y=0]$. Save the outcomes in the corresponding lane in the Rho array, $R[x, 0]$. The processing of each lane in $A[x, y]$, involves three clock cycles: The initial cycle is dedicated to initializing the XOR gate, the subsequent cycle is for computing the XOR functions, and the final cycle is for storing the outputs in the Rho array. Repeat this procedure for the remaining lanes in $A[x, y]$, i.e., $R[x, y]=X\oplus A[x, y]$ where $y \in \{1, 2, 3, 4\}$. The described step is executed over 15 clock cycles for each sheet of $x$.

4. Initialize *lanes* of $R[x+1, y]$ for calculating $\theta(A[x+1, y])$.

In a subsequent section, we extend the circuit depicted in Fig. 8 and save the results of step 3 adjacent to the state array $A$.

### 4.3 Implementation of Operation Pi

The circuit illustrated in Fig. 8 is capable of implementing the Pi operation. This operation involves mapping lanes of $R$, which store $\rho(A)$, back to $A$, as clarified by equation (3). This mapping is facilitated through the routing sheet on the left and CMOS transmission gates on the right-bottom. The execution of this step takes 25 clock cycles.

### 4.4 Implementation of Operation Chi

To implement this operation, it is necessary to expand the state array $A$ according to the configuration shown in Fig. 9. On the right side, each slice not only retains 25-bit state values but also includes their complements.

The Chi array is set up to perform a sequence of two-level AND-XOR logic functions (Fig. 9). Specifically, each wired-OR gate in the Chi array, represented by horizontal wires, is connected to a 2-input volistor XNOR gate, computing an AND-XOR logic function. In each row of the state array (A and ¬A), inputs are fed into five AND-XOR logic functions consecutively. Each function is executed along a designated horizontal wire in the Chi array within a single clock cycle. Therefore, this computational step extents five clock cycles. The outputs of the AND-XOR functions are temporarily stored in the corresponding XNOR gates. Within the next two clock cycles, the input row in the state array will be initialized to permanently store the outputs of the XNOR gates, completing the process within one clock cycle. In this configuration, consider the upper wire in this array executing the operation ¬$A[4, 2] \wedge A[0, 2]$. Here, the memristors associated with the gate's inputs are set to the LRS, while the non-participating memristors are connected to 0V, similar to the configuration shown in Fig. 2c. It's crucial to emphasize that the applied values are the negated inputs. As a result, the output of the wired-OR gate is (¬$A[4, 2] \wedge A[0, 2]$), and this output is directed into the XNOR gate, performing the computation $\chi(A[3, 2]) = A[3, 2] \oplus (\neg A[4, 2] \wedge A[0, 2])$. The output of the XNOR gate, $\chi(A[3, 2])$, is temporarily retained within the XNOR gate. Over the subsequent clock cycles, this output will be permanently stored in $A[3, 2]$.

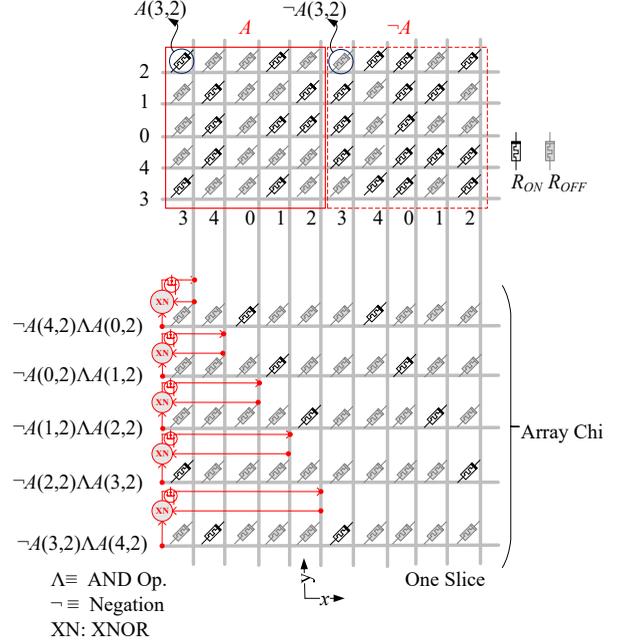

Fig. 9. Implementation of the Chi operation.

Note that the Chi operation is applied to plane $y$, which consists of 64 parallel rows corresponding to the state array. In each of these rows, a combination of five AND-XOR logic gates is implemented within *nine* clock cycles. The entire process is repeated for $y$ values of 0, 1, 2, 3, and 4. During the implementation of $\chi(A)$, the memristors of the wired-OR gates maintain their resistance states. Therefore, there is no need to re-program these memristors for the subsequent Chi operation implementation. The steps for computing the Chi operation are outlined below, and these steps are executed over a span of 50 clock cycles.

1. Calculate the complementary values for the state array $A$ and save the outcomes in the array ¬$A$. The complement of each element $a_{ij} \in A$ is denoted as ¬$a_{ij} \in \neg A$, where $i$ and $j \in \{0, 1, 2, 3, 4\}$. This particular operation is executed within *five* clock cycles.

*For $i$ = 0 to 4 {*

2. Initialize the XNOR gates.
3. Execute the Chi operation on plane $y = i$ within the

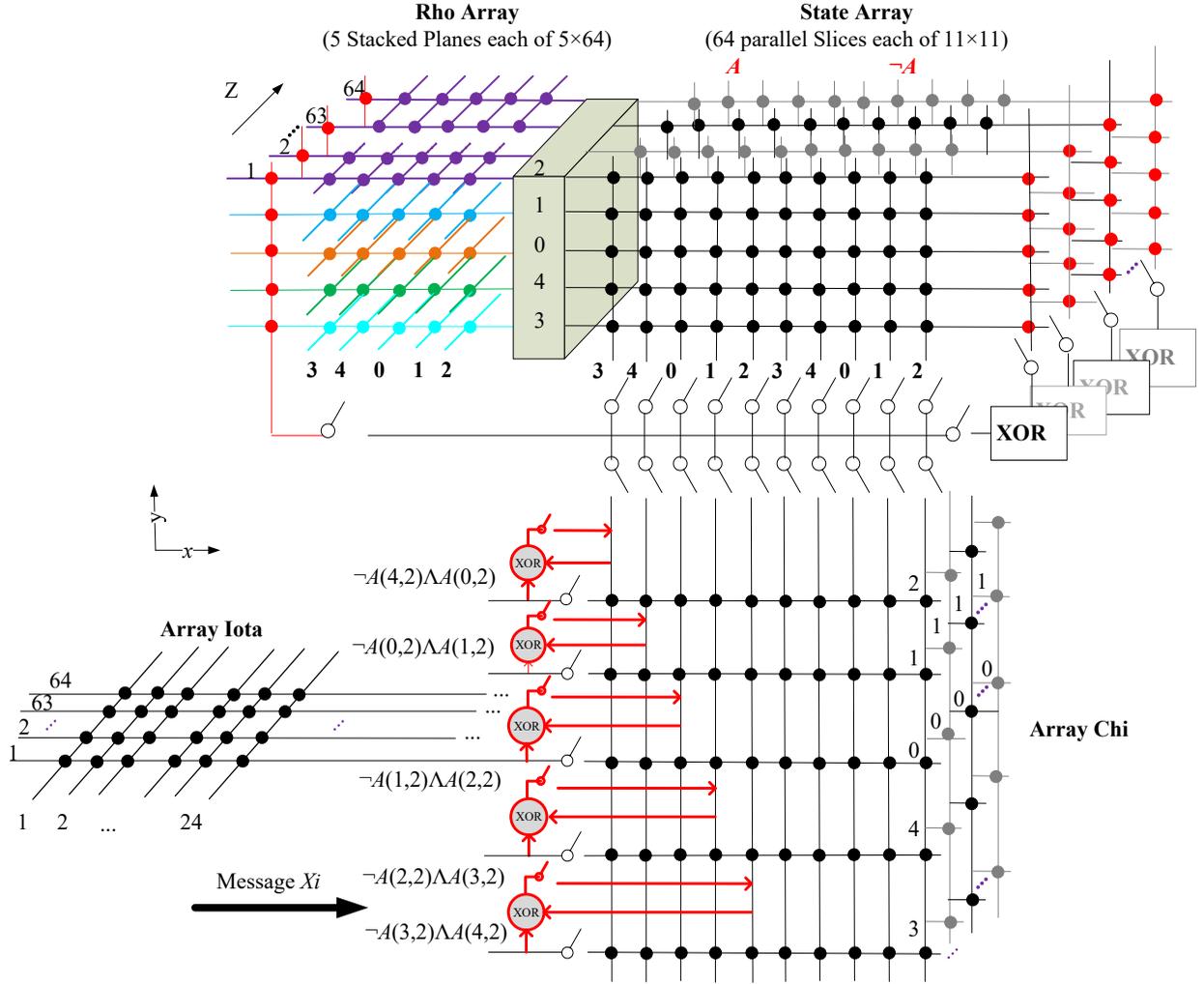

Fig. 10. Proposed accelerator for performing the SHA3 standard. Input voltages (message $X_i$) are applied to the horizontal wires of the Chi array, and the outputs are the final states of state array A.

   state array *A*, and save the outcomes in the respective 2-input XNOR gates.
4. Initialize plane *y=i*.
5. Store the outputs of the XNOR gates in plane *y=i*.
   }
*End*

### 4.5 Implementation of Operation Iota

This implementation requires establishing a connection between the Iota and Chi arrays. Specifically, the Iota array is associated with plane *y*=0 of the Chi array, and only the XOR gates linked to this plane are utilized for the execution of the Iota operation. Fig. 10 illustrates the circuit schematic implementing the Iota operation, along with all preceding operations in the Keccak-*f* function. Within the Iota array, there are 24 constants denoted as *RC[i]*, each consisting of 64 bits. These constants are employed in the bitwise XOR operation $RC[i] \oplus A[0, 0]$ during each round of the Keccak-*f* function. This particular step is carried out over a period of five clock cycles per round.

TABLE 2
VOLTAGE LEVELS APPLIED TO CROSSBAR ARRAYS

| Arrays | Voltage Levels |
|---|---|
| $A$, $\neg A$ | $0, v^+, 2v^+$ |
| Rho | $0, v^+, 2v^+$ |
| Chi | $0, v^+$ |
| Iota | $0, v^+$ |

In our simulations, $v^+$ =0.6V, $V_{SET}$ =1.2V, and $V_{CLEAR}$ =-1.2V.

### 4.6 Input Mapping into the State Array

The computation of $X_i \oplus r$ is essential before executing θ, ρ, π, χ and ι (Fig. 3). This computation involves applying $X_i$ and *r*, signified as $A[x, y]$, to the horizontal and vertical wires of the Chi array, respectively. XOR gates perform the computation and save the results. Subsequently, these results are transferred to the planes of the state array *A*. This programming procedure requires four clock cycles per plane, with the initial cycle dedicated to bitwise XOR operations, the second cycle for plane initialization, the third cycle for storing the output values of the XOR gates in the plane, and the fourth cycle for initializing the XOR gates.

At the beginning, the values in the state array *A* (represented as *r*) are initialized to logic 0. This indicates that the expression $X_i \oplus r$, which simplifies to $X_i$, represents the process of mapping $X_i$ into the state array *A*. Once the input $X_i$ is mapped into the state array, the operations θ, ρ, π, χ and ι will be carried out as explained.

### 4.7 CMOS Drivers

In this study, we implemented a basic design for the CMOS drivers, aiming to illustrate how structuring the crossbar arrays can impact logic and circular bit rotation operations. A wire in a crossbar array connects to a defined voltage level through a transmission gate, with a control bit stored externally. As a result, a wire that needs connections to various voltage levels (e.g., 0, $v^+$, and $2v^+$) in different clock cycles is attached to parallel transmission gates, each corresponding to a specific voltage level. It is important to note that the data organized within the crossbars yields a CMOS peripheral circuitry of minimal size. Every set of crossbar arrays, like the state arrays, functions concurrently and performs identical operations in each clock cycle. Consequently, identical rows or columns within the state arrays and other arrays must receive the same voltage levels, given their execution of identical operations. As a result, a single CMOS driver is employed to control identical rows or columns within the arrays. The arrangement of the crossbar arrays, as employed, is clearly effective in minimizing the CMOS periphery.

## 5 IMPLEMENTATION ANALYSIS

In this section, we analyzed the performance and area overhead of the circuit shown in Fig. 10. Additionally, we evaluated the functionality and energy efficiency of the circuit through simulation.

**Simulation setup**: The simulations were carried out using the LTspice simulator, employing a rectifying memristor model [39] and utilizing voltage pulses with a pulse width of 1 ns and amplitudes specified in TABLE 2. The switching delay, which is the time taken to program a memristor from a HRS or $R_{OFF}$ to a LRS or $R_{ON}$, was considered to be 1ns. This estimation takes into account the conservatively scaled characteristics of memristors achievable for fabrication in the near future [5], [40], [41]. An analogous assumption was applied in the state-of-the-art designs outlined in TABLE 4 for the purpose of comparison. Additional memristor parameters, like $R_{ON}$ and $R_{OFF}$, were set at 500KΩ and 500MΩ, respectively, in accordance with [30].

**Performance Analysis**: Fig. 11 provides an overview of the circuit analysis for a single round of the Keccak-*f* function, including the computational delay at each step. The inputs to the circuit are message blocks $X_i$, which are applied to the XOR gates of the Chi array. The outputs of the circuit correspond to the final states of the memristors in the crossbar array *A*. The hashing of a 1088-bit message block incurs a computational delay of 6326 clock cycles. This includes the 24 rounds, each requiring 263 clock cycles, along with an additional 14 clock cycles. The additional cycles are allocated for input mapping, specifically for computing $X_i \oplus r$, and subsequently storing the results in array *A*. Fig. 11 shows the number of clock cycles per operation, with the total sum of clock cycles for all operations being 263 per round. The voltage levels applied to each crossbar array are detailed in TABLE 2. These values are chosen to satisfy the condition specified in equation (6).

| | |
|---|---|
| 0. | Initialize the state array and the Rho array (2 clock cycles) |
| 1. | *i* = 0; |
| 2. | *while* (*i* <= *t*-1) |
| 3. | Apply *Xi* to array Chi for calculating $X_i \oplus r$ and store the results in array *A* (3 clock cycles per block size *r* ≤320)† |
| 4. | *for j* = 1 to 24, |
| 5. | Implement Theta and store the output in array ¬*A*, (175 clock cycles) |
| 6. | Initialize the Rho array, (1 clock cycle) |
| 7. | Implement Rho operation and store the outputs in the Rho array, (5 clock cycles) |
| 8. | Initialize arrays *A* and ¬*A*, (2 clock cycle) |
| 9. | Implement Pi operation and store the results in array ¬*A*, (25 clock cycles) |
| 10. | Compute complement of ¬*A* and store the results in *A*, (5 clock cycles) |
| 11. | Implement Chi operation and update the values of *A*, (45 clock cycles) |
| 12. | Implement Iota operation and update values of *A*[0, 0], (5 clock cycles) |
| 13. | *end* |
| 14. | *i*++; |
| 15. | *end while* |

Fig. 11. Steps for Implementing the Keccak-*f* function. *t* is the number of message blocks. †When *r* = 1088 or *r* = 576, the mapping delay is 12 or 6 clock cycles, respectively.

$$\begin{cases} v^- - v^+ = V_{CLEAR} \\ V_{SET} = 2v^+ \\ V_{CLEAR} = 2v^- \end{cases} \quad (6)$$

**Area Overhead**: The proposed architecture consists of crossbar arrays, MUXs, and CMOS peripheral circuits. The size of the CMOS peripheral circuitry in our design is kept to a minimum. Each set of crossbar arrays (such as the state arrays, Rho arrays, etc.) operates in parallel and executes the same operation in every clock cycle. Consequently, identical rows or columns in the state arrays, as well as other arrays, must be subjected to the same voltage levels since they carry out identical operations. Therefore, a single CMOS driver is utilized to drive identical rows or columns in the arrays. As an example, a single CMOS driver is employed to drive identical columns across all 64 slices within the state array, effectively reducing the size of the peripheral circuitry.

The CMOS drivers utilize 140.754 KB of control bits, stored in an external memory. For example, 2314 control bits are allocated for mapping a 1088-bit message block into the state array. This allocation is calculated by multiplying the number of control bits linked to the CMOS periphery by the necessary number of clock cycles for the mapping process. In our design, the CMOS drivers for the state array, Rho array, Chi array, and Iota array use 51, 61, 5, and 25 control bits, respectively. Furthermore, an additional 36 control bits are utilized for other CMOS switches, including those that connect the crossbar arrays and control select bits of the MUXs. As a result, the calculation of the 2314 bits is determined by multiplying the total of 178 control bits (covering all previously mentioned control bits) by 13 clock cycles required for the mapping of the message block (Fig. 11).

TABLE 3
ENERGY CONSUMPTION IN THE FIRST ROUND OF HASHING THE PLAIN MESSAGE (PJ)

| Operations | Simulated Sequence of Operations | Energy Consumption |
|---|---|---|
| Initialization | 1. Programming the state array ($A$ and $\neg A$) to HRS and subsequently to LRS | 0.716 |
|  | 2. Programming the Rho array to HRS | 0.004 |
| Mapping the message block (Sec 4.6) | 1. Calculating the XOR gates | 7.659 |
|  | 2. Mapping the message block in the state array $A$ |  |
|  | 3. Initializing the XOR gates |  |
| Performing the Theta operation (Sec 4.2) | 1. Performing $\rho(A[],1)$ | 0.24 |
|  | 2. Computing 10-input XORs | 46.002 |
|  | 3. Computing $X \oplus A[x, y]$ lane by lane |  |
|  | 4. Mapping the outputs in $\neg A[x, y]$ lane by lane |  |
| Performing the Rho operation (Sec. 4.1) | 1. Programming the Rho array to HRS | 0.004 |
|  | 2. Performing Rho | 0.024 |
| Performing the Pi operation (Sec 4.3) | 1. Programming the state array ($A$ and $\neg A$) to HRS and subsequently to LRS | 0.736 |
|  | 2. Performing Pi |  |
| Performing the Chi operation (Sec 4.4) | 1. Computing the negation of $A$ | 0.031 |
|  | 2. Initializing 2-input XOR gates | 8.731 |
|  | 3. Computing 5 {AND-XOR} logic functions |  |
|  | 4. Programming the state array to HRS and subsequently to LRS |  |
|  | 5. Mapping the output results in $A$ |  |
|  | 6. Initializing the XOR gates |  |
| Performing the Iota operation (Sec 4.5) | 1. Calculating the XOR gates | 0.034 |
|  | 2. Initializing $A[0, 0]$ to HRS and subsequently to LRS |  |
|  | 3. Mapping the resulting XORs to $A[0, 0]$ |  |
| Overall energy consumption in crossbar memristors and XOR gates |  | 64.181 |

The crossbar arrays, comprising arrays $A$, $\neg A$, Rho, Chi, and Iota, have a combined size of 1.192 KB. Additionally, the total size of the XOR/XNOR gates and routing sheets is determined to be 0.304 KB. Consequently, the overall size of the memristor circuit is 1.496KB.

**Energy**: The energy consumption in the first round was calculated using the LTspice simulator, encompassing the initialization of memristors, the mapping of inputs, and the execution of round operations. The simulated sequence of operations and their corresponding energy consumptions are presented in TABLE 3. Additionally, the components of the circuit engaged in each operation are outlined below for clarity.

The initialization of the crossbar arrays includes setting the state array (both $A$ and $\neg A$) to the LRS and resetting the Rho array to the HRS. The state array is programmed to the LRS by first resetting the crossbars to the HRS or $R_{OFF}$

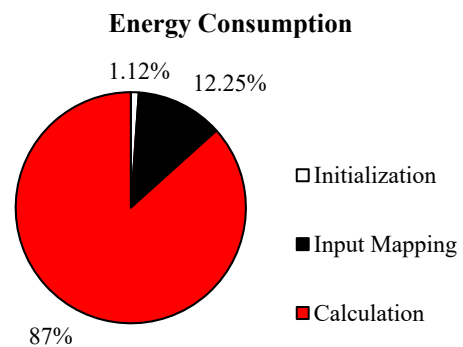

Fig. 12. Energy consumption. Energy consumption in memristor arrays and XOR gates is about 65 pJ per hashing a message block.

TABLE 4
HARDWARE IMPLEMENTATION RESULTS OF SHA3

| Implementation | Freq (MHz) | Size | | CMOS (KGE[2]) | Latency (Cycles) | Throughput (Gbps) | Energy (μJ) |
|---|---|---|---|---|---|---|---|
| | | Memristor[1] (KB) | | | | | |
| | | Instruction | Computing | | | | |
| 65nm ASIC [42] | 1K | - | - | 105 | 12 | 48 | - |
| 90nm ASIC [43] | 455 | - | - | 10.5 | 25 | 19.32 | - |
| 130nm ASIC [44] | 0.1 | - | - | 47.63 | 24 | $4.533 \times 10^{-2}$ | - |
| HashPIM [7][3] | 333 | - | 1048 | - | 3.494K | 39.2 | $0.29 \times 10^{-3}$ |
| ReVAMP [8][4] | 1K | 80 | 0.336 | - | 27.92K | $3.897 \times 10^{-2}$ | 1.53 |
| VG-MTJ [10][5] (Single message) | 402 | - | - | 147.74 | 10.993K | $3.975 \times 10^{-2}$ | 0.39 |
| SHINE-1 [11] | 2K | - | 50.944 | 25.4 | 264 | 33.4 | $4.13 \times 10^{-3}$ |
| **This Work** | **1K** | **140.754** | **1.496** | **5.6** | **6.326K** | **$17.27 \times 10^{-2}$** | **$0.072 \times 10^{-3}$** |

These data were obtained for hashing one message block (i.e. 24 rounds of function $f$ in Fig. 5).

[1] Instruction memristors are used as ROM.

[2] The area is given in terms of gate equivalent (GE). 1 GE is the area of a minimum-sized NAND gate.

[3] The design consists of 378 computational units where the amount of energy consumption per unit is 0.765nJ.

[4] The energy was calculated by the authors of [10] for comparison reasons.

[5] The authors of [10] used crossbar arrays of voltage-gated Spin Half Effect (SHE) Driven magnetic tunnel junction (MTJ) for computing SHA3 where the size of the proposed circuit is 0.361 mm² (or equivalently 147.74KGE in 65nm technology node). The authors also designed a Multiple Message Hash that holds 5 block messages simultaneously.

and then setting them to the LRS or $R_{ON}$. This additional step in the programing of the state array aims to significantly reduce energy consumption in memristors. The initialization step involves energy consumption confined to the state array and the Rho array, with the remainder of the circuit remaining inactive.

In the input mapping step, an arbitrary message block is mapped into the state array $A[x, y]$. This process, detailed in Section 4.6, involves energy consumption limited to the state array, the Chi array, and the 2-input XOR gates connected to the Chi array. At the same time, the remaining part of the circuit remains in an idle (or power-gated) mode.

In the final step, the round operations are executed in the specified order of θ, ρ, π, χ, and ι. The simulated sequence of operations for each round is outlined in TABLE 3. Depending on the specific round operation, only the relevant components of the circuit are active, while the rest of the circuit remains idle. The components that are active for each operation are specified below:
- Theta operation: State array, MUXs, Rho array, and multi-input XOR gates
- Rho operation: State array, MUXs, and Rho array
- Pi operation: Rho array and state array
- Chi operation: State array, Chi array, and 2-input XOR gates
- Iota operation: state array, Chi array, corresponding 2-input XOR gates, and Iota array

The total energy consumption for memristor arrays and XOR gates in the first round is about 65 pJ. Detailed energy breakdown for each step is provided in TABLE 3. Fig. 12 shows the energy consumption in crossbar arrays and XOR/XNOR gates across various stages: the initialization of the state array and the Rho array, the mapping of the message block $X_i$ in crossbar $A$, and computations, which represent the most energy-intensive operation. It's noteworthy that 62% of the energy expended during computations is attributed to the initialization of memristors.

**Comparisons**: An extensive comparison of different architectures is beyond the scope of our paper. However, to get some level of comparison, here are some published numbers for other implementations of SHA3 using both memristive and CMOS technologies. The direct comparison between CMOS and non-volatile memory (NVM) implementations is detailed in TABLE 4, covering energy, throughput, and area. Throughput is computed based on equation (7), and the size of a message block in all implementations is consistent at 1088 bits.

$$Throughput = \frac{Blocksize}{Latency} \times Frequency \quad (7)$$

In TABLE 4, the first three rows present examples of SHA3 implementations utilizing 65nm, 90nm, and 130nm CMOS technologies [42], [43], [44]. Subsequent rows display NVM-based implementations of SHA3. Concerning area, our accelerator exhibits a small area overhead compared to NVM-based implementations.

In the context of energy consumption, our accelerator, encompassing crossbar arrays, volistor XOR gates, and

MUXs, surpasses NVM-based implementations by consuming 72 pJ. It's important to note that MUXs are active for only 6 out of 6326 clock cycles. Their implementation utilizes a 65nm standard cell library and incurs a consumption of 7.6 pJ. In comparison to recent CMOS technology, our accelerator exhibits lower energy consumption. Specifically, the energy consumed in the datapath for accessing a single message block (1088 bits) from an HBM2 (High Bandwidth Memory) implemented in a 22nm technology node is measured at 3.786 nJ (equivalent to 3.45 pJ/bit [45]). Notably, this energy usage is nearly two orders of magnitude higher than the overall energy consumption in our accelerator. The energy savings in our implementation are attributed to in-memory computations, rendering the mentioned energy unnecessary.

In terms of delay and throughput, our accelerator surpasses both VG-MTJ and ReVAMP. However, it's important to note that SHINE and HashPIM exhibit higher performance than our accelerator, albeit at the expense of larger areas and increased energy consumption. As such, our design is tailored for optimal area utilization and energy efficiency. In summary,

1. The primary difference between our design and previous ones, centers around architectural considerations. This encompasses factors like the size and structure of crossbar arrays, as well as specific attributes of memristors (e.g., rectifying vs. non-rectifying) in the respective designs.
2. In our approach, the Rho operation employs CMOS MUXs connecting perpendicular crossbar arrays. This differs from other methods, where CMOS technology was used for the Rho operation. Notably, [7] handles this operation within the crossbar arrays themselves.
3. Our methodology implements logic functions using both non-stateful and stateful logic gates—programmable diode gates and volistors within crossbar arrays of rectifying memristors. Conversely, alternative techniques employ various logic implementations, including stateful logic gates such as MAGIC [46].
4. Our architecture is tailored to optimize both area and energy concurrently, emphasizing parallel computations. In contrast, alternative approaches, like [7], prioritize performance over area, while [8] prioritizes area optimization at the expense of performance degradation.

# 6 CONCLUSION

Our paper introduces a compact 3D architecture for in-memory computing with SHA3. Using CMOS-memristor technology, our design implements data representation in SHA3 directly. Leveraging memristor characteristics, our design addresses the memory access bottleneck and eliminates standby power in conventional computers. Additionally, our optimized design minimizes both area and energy consumption, highlighting the potential of hybrid CMOS-memristor technology for security implementations with reduced size and energy requirements.